# Real-Time ECG Interval Monitoring Using a Fully Disposable Wireless Patch Sensor

Gabriel Nallathambi
*VitalConnect Inc.,*
San Jose, CA 95110 USA,
gnallathambi@vitalconnect.com

Paurakh Rajbhandary
*VitalConnect Inc.,*
San Jose, CA 95110 USA,
prajbhandary@vitalconnect.com

Thang Tran
*VitalConnect Inc.,*
San Jose, CA 95110 USA,
ttran@vitalconnect.com

Olivier Colliou
*VitalConnect Inc.,*
San Jose, CA 95110 USA,
ocolliou@vitalconnect.com

*Abstract*—ECG interval monitoring provides key insights into the diagnosis of cardiac diseases. The standard 12-lead ECG is generally used, however, because of the current COVID-19 pandemic there is a strong need for a remote monitoring solution which will reduce exposure of health care providers to coronavirus. This article presents a disposable wireless patch biosensor (VitalPatch) and associated platform functionalities for real-time continuous measurement of clinically relevant ECG intervals including PR interval, QRS duration, QT interval, corrected QT interval by Bazett (QTb), and corrected QT interval by Fridericia (QTf). The performance of the VitalPatch is validated by comparing its automated algorithm interval measurements to the manually annotated global intervals of the 12-lead ECG device in 30 subjects. The accuracy of interval monitoring (in terms of mean timing error calculated by subtracting the VitalPatch measurements from the global intervals) is 2.7±15.94 ms, -1.97±12.29 ms, -14.6±12.97 ms, -15.33±14.11 ms, and -15.08±13.69 ms for PR interval, QRS duration, QT interval, QTb, and QTf, respectively. These results demonstrate that the VitalPatch is a viable solution for measuring ECG intervals while taking advantage of its remote monitoring feature during the pandemic.

## I. Introduction

Characterization of electrocardiogram (ECG) intervals is essential in the diagnosis of cardiac diseases. Amongst ECG intervals, monitoring of PR interval, QT interval, and QRS duration is of paramount importance as they are clinically well established indicators of conduction disorders and cardiac disease risk stratification. For instance, strong associations have been established between abnormal PR intervals, QRS durations and QT intervals to the risk of atrial fibrillation [1], sudden cardiac death [2], and ventricular arrhythmia [3], respectively.

The clinical gold standard for interval monitoring is the standard 12-lead ECG device. However, the onslaught of Coronavirus disease 2019 (COVID-19) pandemic has precipitated the need for remote patient monitoring to minimize exposure of health care workers to patients with severe acute respiratory syndrome coronavirus 2 (SARS-CoV-2) virus, and unburden the health care system by reducing the number of patients admitted to the hospital.

Patients with COVID-19 are at risk of rapid clinical deterioration in both the in-patient and out-patient settings. Recent studies have shown anywhere from 7 to 78% of cardiac injuries in COVID-19 patients [4]–[7]. Furthermore, studies suggest that a majority of patients could suffer long-term consequences of COVID-19, and even the mildest cases could cause permanent heart damage [8].

In view of the above evidence, health care systems and regulatory agencies have recognized the critical need for continuous remote monitoring of ECG-based cardiac event markers. In the United States, Food and Drug Administration (FDA) granted VitalPatch® an Emergency Use Authorization (EUA) for remote monitoring of QT intervals during the COVID-19 pandemic [9]. VitalPatch is a fully disposable wireless patch sensor with the capability of measuring ECG, heart rate, heart rate variability, skin temperature, body temperature, respiration rate, step count, body posture, and fall detection [10]–[12]. The utilization of this technology to remotely monitor COVID-19 patients at risk of rapid clinical deterioration virtually eliminates the possibility of spread of infection from a patient to other patients or to caregivers in both the in-patient and out-patient settings, with the added benefit of simultaneous and continuous assessment of vital signs, activity measures, and ECG intervals in a single platform.

In this paper, we first describe the interval monitoring capabilities of the patch biosensor and the associated functionalities of the platform to enable remote patient monitoring. Secondly, we evaluate the performance of VitalPatch biosensor in the automated interval measurement from single lead bipolar ECG compared to simultaneously acquired manually annotated global intervals of 12-lead ECG device.

## II. Materials and Methods

### A. Vista Solution platform and ECG interval measurement

Vista Solution™ is a fully integrated system that allows clinicians to remotely monitor continuous vital signs and ECG interval measurements. Real-time clinical monitoring is enabled by VitalPatch® biosensor that measures and transmits physiological signals via Bluetooth low energy (BLE) to tablet (VistaTablet™) or phone (VistaPhone) relay device. The relay device consolidates and transmits the physiological data to VitalCloud over WiFi or cellular network, and a central workstation VistaCenter web application (shown in Fig. 1a) allows clinicians to monitor multiple patients simultaneously.

Interval measurements are made possible by the proprietary system-on-chip processor that performs real-time computations on single-lead ECG data. The interval measurement algorithms operate as follows: First, the ECG data during a processing window is accumulated and the signal is analyzed in the time domain to detect the presence or absence of noise. Based on the ECG signal quality and the level of noise, beats are detected and the peak and boundaries of QRS complex is identified. QRS analyzer discriminates the types of heart beats based on the shape of waveforms, and average QRS analyzer determines positions of fiducial point of PQRST waveforms and performs measurement of the intervals using statistical analysis methods.

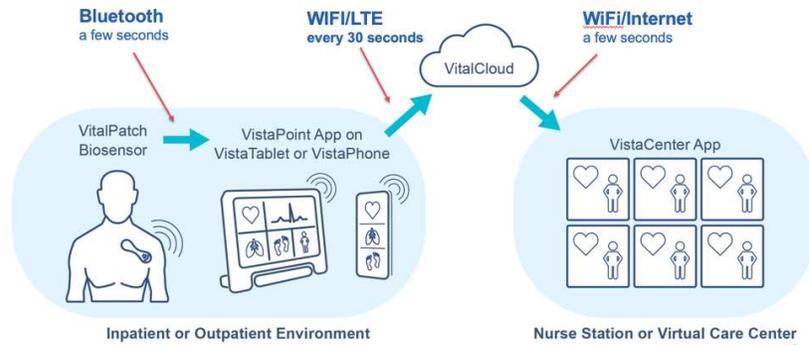

(a)

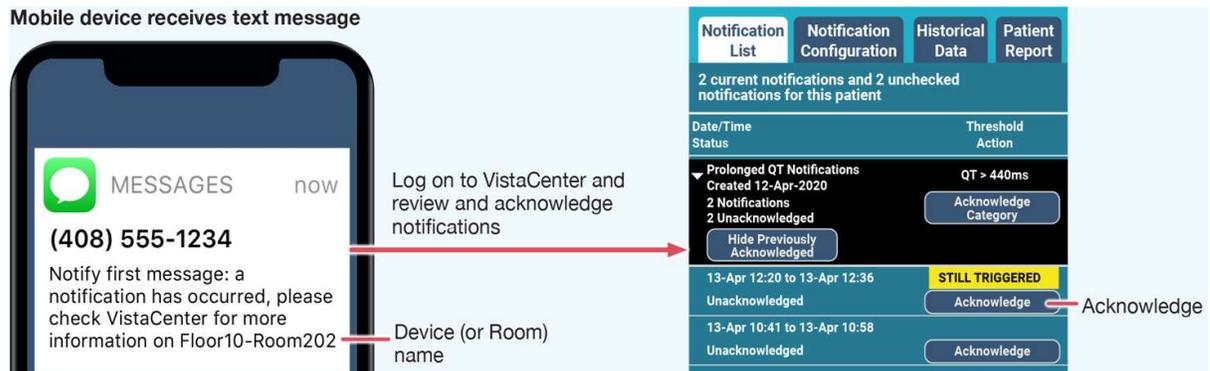

(b)

Fig. 1. (a) Components of the Vista Solution platform for real–time continuous monitoring of ECG intervals. (b) Clinical notification and acknowledgement of interval measurements (Source: VistaCenter 3.0 Instructions for Use [13]).

VistaCenter offers clinicians an option to set automated notifications based on threshold crossing of window averaged ECG intervals [13]. When a notification triggers, healthcare providers can use mobile device to view and acknowledge the notification as shown in Fig. 1b. VistaCenter displays details on a patient's notifications, including time stamps, recent history, and a method to acknowledge notifications. Healthcare experts can review the ECG and verify the automated intervals measured by our technology, and follow up with patients as needed.

### B. Clinical Study Design

Thirty volunteers (n=30, 41±13 years, M/F: 15/15, BMI: 26.2±9.8 kg/m$^2$) participated in an IRB approved testing protocol with informed consent. The clinical protocol utilizes the patch biosensor and a 12-lead Holter monitor to simultaneously acquire data from the subject in stationary position using the electrode placement shown in Fig. 2.

The VitalPatch biosensor is placed on the left chest of each subject after skin preparation with an alcohol wipe. The biosensor is connected to the tablet relay device, and all data including automated ECG interval measurements of the algorithm is wirelessly uploaded to the cloud.

The 12-lead Holter monitor (Cardiocard B12 10-electrode Nassiff Holter system) is placed in the standard Mason-Likar setup (in which limb electrodes are placed in the corners of the trunk of body). The placement location of electrodes is strictly followed per clinical guidelines and adhesives are used to prevent tugging of wires. All ECG data from the 12-lead device is transferred and stored for further analysis.

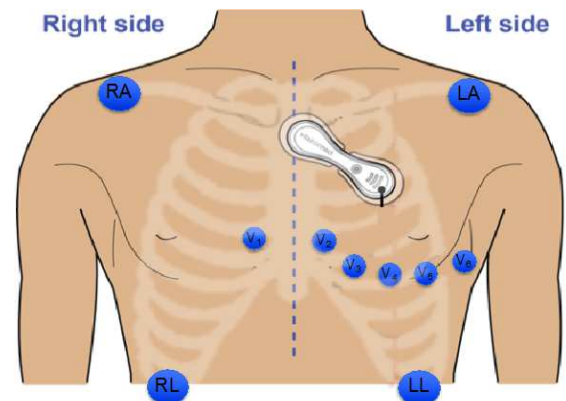

Fig. 2. Electrode placement layout of Holter Monitor (blue electrodes are for Mason-Likar 12 lead ECG system) and VitalPatch.

TABLE I. GLOBAL INTERVAL DISTRIBUTIONS (N=30).

| Characteristics | Percentile | | | | | | | | |
|---|---|---|---|---|---|---|---|---|---|
| | 2.5 | 5 | 10 | 25 | 50 | 75 | 90 | 95 | 97.5 |
| Age (years) | 24 | 25 | 27 | 31 | 36 | 49 | 54 | 60 | 62 |
| Height (inches) | 61.4 | 62.0 | 62.2 | 65.0 | 67.1 | 71.0 | 73.1 | 74.0 | 76.3 |
| Weight (lbs) | 112.5 | 120.0 | 120.0 | 134.5 | 167.0 | 195.0 | 225.0 | 246.0 | 309.0 |
| BMI (kg/m2) | 18.6 | 19.4 | 20.0 | 22.0 | 24.1 | 28.0 | 35.1 | 40.4 | 47.3 |
| PR interval (ms) | 104.0 | 112.0 | 120.0 | 136.0 | 144.0 | 160.0 | 184.0 | 192.0 | 216.0 |
| QRS duration (ms) | 88.0 | 88.0 | 96.0 | 96.0 | 104.0 | 112.0 | 120.0 | 128.0 | 128.0 |
| QT interval (ms) | 320.0 | 336.0 | 344.0 | 368.0 | 392.0 | 408.0 | 432.0 | 440.0 | 456.0 |
| QTb interval (ms) | 358.2 | 364.0 | 370.1 | 386.7 | 403.5 | 429.1 | 452.0 | 468.8 | 482.2 |
| QTf interval (ms) | 353.1 | 359.0 | 366.0 | 380.6 | 398.7 | 419.5 | 438.7 | 447.8 | 456.1 |
| P duration (ms) | 80.0 | 88.0 | 88.0 | 96.0 | 104.0 | 120.0 | 128.0 | 136.0 | 136.0 |

*C. Annotation of Global Intervals and Duration*

The ECG data from the 12-lead Holter monitor is carefully visually assessed for signal quality, and one-minute ECG segment having high signal-to-noise ratio is selected for annotation of global intervals. The global annotations are based on the earliest onset or the latest offset ECG point of a morphology under consideration in the twelve ECG channels per the FDA recognized consensus standard [14]. Global annotations for the intervals and duration of interest is defined as follows:

- PR interval: Earliest onset of QRS complex - Earliest onset of P wave.
- QRS duration: (Latest offset - Earliest onset) of QRS complex
- QT interval: Latest offset of T wave - Earliest onset of QRS complex

Additionally, corrected QT intervals are computed from the global QT interval and RR interval as follows:

- Corrected QT interval by Bazett (QTb): $\frac{QT}{\sqrt{RR}}$
- Corrected QT interval by Fridericia (QTf): $\frac{QT}{\sqrt[3]{RR}}$.

*D. Performance Validation*

The ECG interval measurement algorithm of the patch biosensor is validated by comparing the automated intervals measured by VitalPatch against the manually annotated global intervals. The accuracy is evaluated in terms of the timing error ($t_e$) and mean timing error ($t_{me}$) calculated by subtracting the VitalPatch measurements from the global intervals.

III. RESULTS

Table I shows the distribution of annotated global interval parameters spans a wide range with the median values for PR, QRS, QT, QTb, and QTf being 144 ms, 104 ms, 392 ms, 403.5 ms, and 398.7 ms, respectively.

The results of timing accuracy of interval measurements detected by the algorithm in VitalPatch compared to manually annotated global intervals is provided in Table II. With an ECG sampling frequency of 125 Hz, $t_e$ and $t_{me}$ of automated interval measurement is in the vicinity of 2–sample deviation for PR interval and QRS duration, and 4–sample deviation for QT and corrected QT intervals.

IV. DISCUSSION

Since the emergence of COVID-19 as a global pandemic, there has been an increasing need for devices that can remotely monitor patients in both in-patient and out-patient settings. This article presents a fully disposable wireless patch for continuous real-time ECG interval monitoring and demonstrates clinically acceptable timing accuracy compared to manually annotated global intervals. While automated interval monitoring based on a single-lead patch ECG device will not replace interval measurements from diagnostic 12-lead ECG in a clinical environment, the results establish the feasibility of remote monitoring of ECG intervals using small patch-based devices.

In this article, the performance of interval measurement algorithms is established by comparison against global intervals. Per the AHA scientific statement on standardization and interpretation of the ECG [15], the use of simultaneous 12-lead ECG in this study enables precise identification of the earliest onset and latest offset in the leads for annotation of global intervals.

Comparative studies based on commercial 12-lead ECG devices have shown that degree of abnormality in the ECG impacts performance of interval measurement algorithms [16]. While prior medical history and abnormality in the ECG rhythm is not separately assessed in this study, it should be noted that the distribution of the annotated global intervals spans both normal and abnormal ranges.

TABLE II. PERFORMANCE OF AUTOMATED INTERVAL MEASUREMENT COMPARED TO GLOBAL INTERVALS

| Interval | $t_{me}$ (ms) | $t_e$ (ms) |
|---|---|---|
| PR interval (ms) | 2.7 ± 15.94 | 2.01 ± 15.02 |
| QRS duration (ms) | -1.97 ± 12.29 | -1.46 ± 13.92 |
| QT interval (ms) | -14.6 ± 12.97 | -15.19 ± 14.83 |
| QTb interval (ms) | -15.33 ± 14.11 | -16.17 ± 16.58 |
| QTf interval (ms) | -15.08 ± 13.69 | -15.77 ± 16.06 |

Other key factors that influence automated interval measurements include patch orientation and sampling frequency. In this study, the patch is placed in a standardized location (one electrode two fingers below the suprasternal notch at a 45-degree angle) to minimize changes in the patch orientation. While measurement variations may arise from minor differences in patch placement, it should not have significant adverse impact on the evaluated performance. On the other hand, choice of sampling frequency inherently limits the timing resolution of the interval measurement. A higher ECG sampling frequency would increase the timing accuracy of the interval measurement; however, it would drastically reduce the battery life of the wireless patch sensor. The tradeoff between timing resolution and battery life will highly depend on the clinical use case scenarios.

Besides continuous monitoring of ECG intervals, the VistaSolution platform also measures key vital signs with configurable real-time notifications enabling healthcare providers to remotely monitor patients in both hospital and home settings. Previous clinical studies have well established the effectiveness of the patch sensor technology in hospital general wards [17], hospital step down units [18], and home use [19] for remote monitoring of patient vital signs and actigraphy. As VitalPatch is a fully disposable single-use device with battery life of 7 days, it enables long-term monitoring, eliminates cross-contamination, and provides unparalleled convenience to clinicians.

In conclusion, the VitalPatch has been demonstrated to be a viable remote monitoring solution for measuring ECG intervals. This will allow health care providers to monitor and diagnose cardiac diseases while minimizing their exposure to COVID-19 or other infectious diseases, and unburden the health care system by reducing the number of patients admitted to the hospital.